\documentclass[journal]{IEEEtran}
\usepackage{amsfonts}
\usepackage[linesnumbered,ruled]{algorithm2e}
\usepackage{algpseudocode}
\usepackage{amsfonts}
\usepackage{bm}
\usepackage{amsmath}
\usepackage{exscale}
\usepackage{relsize}
\usepackage{cite}
\usepackage{amssymb}
\usepackage{stfloats}
\usepackage{extarrows}
\usepackage{enumerate}
\usepackage{color,xcolor}
\usepackage{graphicx}
\usepackage{multirow,tabularx}
\usepackage{subfigure}
\usepackage{hyperref}

\usepackage{times,amsmath,color,amssymb,graphicx,geometry, epsfig,psfrag,subfigure}

\newtheorem{theorem} {Theorem}
\newtheorem{remark}[theorem]{Remark}

\newcommand{\tabincell}[2]{\begin{tabular}{@{}#1@{}}#2\end{tabular}}

\makeatletter

\newcommand{\Rmnum}[1]{\expandafter\@slowromancap\romannumeral #1@}
\makeatother

\ifCLASSINFOpdf
\else
\fi
\begin{document}
%


\title{An LSTM-Aided Hybrid Random Access Scheme for 6G  Machine Type Communication  Networks}

\author{\IEEEauthorblockN{Wenchao Zhai, Huimei Han,  Lei Liu,  Jun Zhao}

\thanks{

Wenchao Zhai is with the China Jiliang University, Hangzhou, Zhejiang, 310018, P.R. China, Huimei Han  
is with College of Information Engineering, Zhejiang University of Technology, Hangzhou, Zhejiang, 310032, P.R. China,  Lei Liu is with the State Key Lab of Integrated Services Networks, Xidian University, Xi'an, 710071, P.R. China,  and Jun Zhao is  with School of Computer Science and Engineering, Nanyang Technological University, Singapore. (Email:  zhaiwenchao@cjlu.edu.cn, hmhan1215@zjut.edu.cn tianjiaoliulei@163.com, junzhao@ntu.edu.sg)



}}

\maketitle

\begin{abstract}
In this paper, an LSTM-aided  hybrid random access scheme (LSTMH-RA) is proposed to support diverse quality of service (QoS) requirements in 6G machine-type communication (MTC)  networks, where  massive  MTC  (mMTC)  devices and ultra-reliable  low  latency communications (URLLC)  devices coexist. In the proposed LSTMH-RA scheme, mMTC  devices access the network via  a timing advance (TA)-aided four-step procedure  to meet  massive access requirement,  while  the access procedure of the URLLC devices is completed in two steps  coupled with  the mMTC  devices' access procedure to  reduce latency. 
Furthermore,  we propose an attention-based LSTM prediction model to predict the number of active URLLC devices, thereby  determining   the parameters of the multi-user detection algorithm to  guarantee the latency and reliability access requirements of URLLC devices.
We analyze  the successful access probability of the LSTMH-RA scheme.  Numerical results show  that, compared with the  benchmark schemes, the proposed  LSTMH-RA scheme can significantly improve the successful access probability, and thus satisfy the  diverse QoS requirements of URLLC and mMTC devices.
\end{abstract}


%
\IEEEpeerreviewmaketitle

\section{Introduction}

\IEEEPARstart{W}{ITH}  the rapid development of the Internet of Things (IoT) industry, such as smart cities and autonomous driving, the amount of  data generated by machine-type communication (MTC) accounts for a great proportion in all communication services~\cite{IoT_add,IoT_add1}.
 The fifth-generation (5G) communication standard has defined two types of services for MTC~\cite{5G}: one is massive machine type communications (mMTC), which aims to provide massive connections; the other  is ultra-reliable low latency communications (URLLC),  which focuses on supporting  high reliability and low latency communications. However, the key  performance indicators of 5G are not enough to  meet some future MTC  requirements~\cite{6g}, such as the 5G air interface delay  (less than 1 ms) is difficult to meet the haptic internet-based telemedicine  air interface delay (less than 0.1 ms). To tackle these problems, several research projects around the world have recently studied the next generation mobile communication systems, namely the sixth generation mobile communication systems (6th-Generation, 6G)~\cite{6ge1,6ge3}.

Compared with 5G, the 6G  communication systems will achieve lower power consumption, lower latency, higher reliability, and so on. Thus, 6G will become the main force to support and promote IoT services in the future,  bringing performance benefits and unprecedented services to users~\cite{6ge1}. According to the prediction of Cisco visual networking index, the number of MTC devices will  be around 28.5 billion by 2022~\cite{crowdue}.
 MTC will be a critical cornerstone for the future 6G systems due to the requirement of vertical-specific wireless network solutions. Furthermore,
many MTC applications in 6G  will encompass  both URLLC and mMTC services, whereas  mMTC and URLLC have been often studied separately in 5G~\cite{MTC}. 

Random access  procedure, initiating the data transmission,   is a critical step  in 6G communication systems~\cite{MTC}.
 The design of  random access schemes  for the MTC network has attracted much attention in recent years. 
The third generation partnership project (3GPP) has  proposed several congestion control approaches to decrease the network load in the random access procedure, thereby improving  the uplink throughput~\cite{3gpp1}. The popular one is access class barring (ACB) based methods where devices whose generated random values are less than or equal to the ACB factor can access the network. Several enhanced ACB-based random access schemes were proposed, such as dynamic-ACB-based
schemes~\cite{ACB1,ACB2},  cooperative-ACB-based scheme~\cite{ACB3}, and joint ACB and timing advance (TA) based 
 scheme~\cite{ACB4}. Moreover, other congestion control strategies, such as slotted/backoff based approaches, were also described in~\cite{3gpp1}. 
 Considering the small data packet transmission of MTC devices, to reduce the signaling overhead,  Laya~\textit{et~al.}~proposed to transmit the data information of MTC devices during MSG1 or MSG3  transmission stage~\cite{data}. The above-mentioned  schemes  are orthogonal random access (ORA) schemes,  where each resource block  can only be allocated to a single MTC device.  This will decrease  data transmission efficiency, reduce resource utilization and the  number of successful devices,  as well as increase the access delay and  energy consumption.

The  non-orthogonal random access (NORA) schemes, which allows multiple MTC devices share the same resource block to overcome the drawbacks of ORA,  attracted much attention in recent years.  Seo~\textit{et~al.}~proposed a power-domain non-orthogonal multiple access based random access scheme, where  multi-channel selection diversity is utilized to improve the energy efficiency~\cite{mul}.  Shirvanimoghaddam~\textit{et~al.}~proposed a  rateless analog fountain code based NORA scheme, where  MTC devices are grouped based on their delays, and devices from  the same group  transmit their uplink messages to the BS via the same resource block~\cite{NORA1}.  Liang~\textit{et~al.}~proposed a power domain based NORA scheme, where MTC devices transmit their uplink messages with different transmitting powers via the same resource block, and the BS recovers the uplink message of each device by employing a successive interference cancellation (SIC) algorithm~\cite{NORA2}.
Considering sparse activity  of MTC devices, the problem of detecting uplink message can be transformed into the problem of sparse signal recovery in compressed sensing. Then, the BS  performs active MTC device detection, channel response estimation, and uplink message decoding by utilizing a  measurement matrix~\cite{ce5,ce8}.  Ahn~\textit{et~al.}~proposed a
Bayesian-based random access scheme for the case of the BS with a single antenna, where an expectation propagation algorithm is utilized to perform activity detection and channel estimation jointly~\cite{ce5}. 
 Liu~\textit{et~al.}~proposed an  approximate message passing (AMP) based grant-free scheme to solve the massive access problem in massive MIMO systems~\cite{ce8}. Machine learning has strong learning and expression capabilities and has been widely used in  communications~\cite{ML1,ML2,ML3}. Researchers conduct research on  machine learning based NORA scheme in the past few years.
 Luis~\textit{et~al.}~proposed a reinforcement learning based  NORA scheme~\cite{ML1}. In this scheme,  ACB parameters are used as reinforcement learning actions,  the number of collision-free preambles is used as the reward value of reinforcement learning, and the ACB parameter values under the  different number of active devices are learned. 
 Gui~\textit{et~al.}~proposed a long short-term memory (LSTM)-based NORA scheme where an LSTM-based  deep learning model is utilized to learn the channel characteristics between the BS and the device for  better  power allocation, and then the  multi-user detection technology is used for  uplink message decoding~\cite{ML2}.
 Ye~\textit{et~al.}~established an end-to-end NORA network model,  where deep variational autoencoder is utilized to realize the  uplink message decoding and active UE detection~\cite{ML3}. However, the above studies only have considered the case of all devices having the same priority, which are not suitable for  MTC network with  diverse quality of service (QoS) in 6G~\cite{MTC}.

 To solve this problem,  Weerasinghe~\textit{et~al.}~proposed a group-based random access scheme, where  URLLC devices are grouped and  some preambles
are preserved for these devices, whereas mMTC devices still utilize the traditional orthogonal access mechanism. However, the traditional orthogonal access mechanism cannot meet the requirements of massive connections of mMTC devices, leading to network congestion~\cite{co1}.  Qi~\textit{et~al.}~proposed a multi-channel ALOHA random access scheme. This scheme divides the channel resources in the time domain and the frequency domain to obtain multiple sub-channel resources, and provides more access resources for URLLC devices to ensure the low-latency and high-reliability connection of  URLLC devices. However, when multiple devices send uplink messages via the same sub-channel, this scheme cannot decode the uplink messages of these devices, and thus limits the number of successful access devices~\cite{co2}.

To guarantee diverse QoS requirements in 6G  MTC networks where URLLC and mMTC devices coexist, we propose an LSTM-aided  hybrid random access (LSTMH-RA) scheme. In the proposed LSTMH-RA scheme, mMTC  devices access the network via  a  TA-aided four-step procedure, while  the access procedure of the URLLC devices is completed in two steps  coupled with  the mMTC  devices' access procedure. 
Furthermore,  we propose an attention-based LSTM prediction model to predict the number of active URLLC devices, thereby  determining   the parameters of the multi-user detection algorithm to  guarantee the latency and reliability access requirements of URLLC devices.
 The \textbf{main contributions} of this paper are summarized as follows\footnote{Since we have also uploaded this paper to arXiv, the arXiv paper \cite{arxivupdate} is the same as this paper.}. 
\begin{enumerate}[1)]


     \item 
     We propose an attention-based LSTM prediction model to predict the number of  active URLLC devices. Thus, the BS can determine the parameters of muti-user detection  dynamically to meet the reliability requirement. Furthermore, this also ensures that almost all URLLC devices  can access the network  in one shot to satisfy the latency requirement.
     
     \item 
     We utilize the SIC algorithm in power domain to decode the uplink message of mMTC devices. {\color{red}The SIC algorithm decodes the uplink messages of mMTC devices from the highest to the lowest power level.
As long as mMTC device with the highest power level experiences power collision, other mMTC devices with/without power collision will not  be successfully decoded. To enable other mMTC devices without power collision have opportunity 
to be decoded successfully, we propose to allocate the highest power level to mMTC devices  experiencing  no TA collision with contenders selecting the same preamble, and allows other mMTC devices to randomly select their power levels.}

     
     

    \item We analyze  the successful access probability of the proposed LSTMH-RA scheme.  
    Numerical results show  that the simulation results accord  well with the analytical  results. In addition, compared with the benchmark schemes, the  proposed scheme significantly improves the successful access probability, and thus satisfies the  diverse QoS requirements of URLLC and mMTC devices.
\end{enumerate}

The remainder  of this paper is organized as follows.
 Sections II and III  introduce the system model and the proposed  LSTMH-RA scheme, respectively. In Section IV, we analyze the performance of the proposed LSTMH-RA scheme. Sections V and VI provide numerical results and  conclusions, respectively.

\textbf{Notation.} Table~\ref{tab:notation} describes the  notations utilized throughout this paper. 
\begin{table}[htbp]
\scriptsize
\centering
 \caption{notations.}\label{tab:notation}
   \begin{tabular}{|c|c|}
\hline
\linespread{2}
Notations & Description\\
\hline
$(\cdot)^T$ &~ \tabincell{c}{ The transpose  operation  of  a \\
 vector or a matrix} \\
 \hline
$|\cdot|$& The cardinality of a set\\
\hline
$\mathrm{concat(\cdot)}$& The concat function\\
\hline
$\mathrm{exp(\cdot)}$& The exponential function \\
\hline
$\mathcal{C}\mathcal{N}( \mu, \sigma^2 )$& ~ \tabincell{c}{A circularly-symmetric complex Gaussian \\ distribution  with mean $\mu$  and variance $\sigma^2$}\\
\hline
$\mathrm{tanh(\cdot)}$& The  hyperbolic tangent function\\
\hline
$| a |$& The modulus of complex number $a$\\
\hline
$\lceil \cdot \rceil$& Round up to an integer\\
\hline
$\mathcal{P}(\lambda)$ & Poisson distribution with parameter $\lambda$\\
\hline
$\mathcal{B}( b,c)$ &  Binomial distribution  with parameters b and c\\
\hline
$\otimes$ & Circular convolution operation\\
\hline
$[\bm{d}]_i$ & The $i^{th}$ element of  vector $\bm{d}$\\
\hline
$\mathbb{C}$& The complex numbers space\\
\hline
\end{tabular}
\end{table}

\section{SYSTEM MODEL }
\begin{figure}[htbp]
	\centering
\includegraphics[scale =0.5] {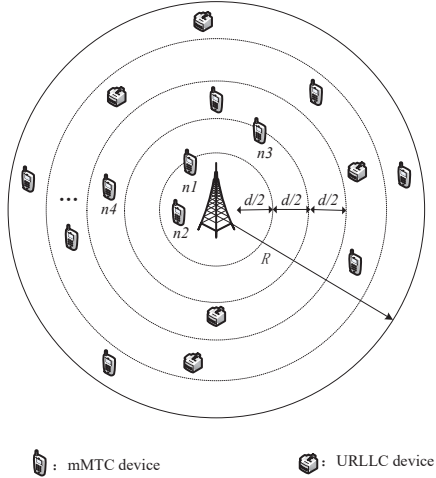}\\
	\caption{System model.}\label{System}
\end{figure}

In this paper, we consider a cell with radius of $R$, where a BS is in the center of the cell and all $K$ devices (i.e., mMTC devices and URLLC devices)  are uniformly distributed in the cell, as shown in Fig.~{\ref{System}}. Furthermore, the BS is equipped with $M$ antennas and each device has a single antenna.
 The number of the active devices is $N_a$, including $N_\text{U}$ URLLC devices and $N_\text{M}$ mMTC devices, and there are $\tau_p$ available preambles. {\color{red} In addition, we consider the block fading channel, and the system is in the time division duplex (TDD) mode. Similar to \cite{SUCR}, the steps of  the proposed LSTMH-RA scheme are performed in a single coherence interval, and the downlink channel can be estimated via the uplink pilots due to the channel reciprocity~\cite{add1_2}.}

We utilize the SIC algorithm in power domain to decode the uplink message of each  mMTC device. To improve the performance of the SIC algorithm, we propose to allocate the highest power level to mMTC devices  experiencing  no TA collision with contenders selecting the same preamble, and allows other mMTC devices to randomly select their power levels. Next, we first introduce the TA information, and then describe the power levels.

The TA information is an index value
 after quantizing its round-trip propagation delay with the
granularity of $d=16T_s\times c$~\cite{xiaojie}. Thus,  the maximum TA value  is $\zeta=\lceil{\dfrac{2R}{d}}\rceil$. Apparently, MTC devices in each annulus obtained by quantizing distance $d/2$ from the center to the edge of the cell, have the same TA information~\cite{xiaojie} as shown in Fig.~{\ref{System}}.
We use  $TA_{i}$ to denote the TA index of devices located at the $i^{th}$ annulus.  For example, as shown in Fig.~{\ref{System}}, TA indexes of mMTC devices n1, n2, n3 and n4  are  $TA_{1}, TA_{1}, TA_{2}$, and $TA_{3}$, respectively.
 Furthermore,  each mMTC device can know its distance to the BS via distance measuring technologies, and thus obtains its TA information~\cite{wang}. For the case of mMTC devices n1, n2, n3 and n4 selecting the same preamble, n3 and n4 are TA collision-free mMTC devices.
 



 
In addition, the SIC  algorithm  decodes the  uplink  message  of each mMTC device  from  the  highest  to  the lowest  power levels. Consider that there are $L$ power levels,
denoted by $\{l_1,\cdots,l_j,\cdots,l_L\}$ satisfying $l_1>\cdots>l_L$, and   $l_j~(j=1,\ldots,L)$ is written as~\cite{pp}
\begin{equation}\label{a2}
\begin{array}{l}
 l_j=\gamma(\gamma+1)^{L-j},
\end{array}
\end{equation}
where $\gamma$ is the target signal to interference plus noise ratio (SINR).

\section{The LSTMH-RA Scheme}


\begin{figure}[htbp]
	\centering
	\includegraphics[scale =0.65] {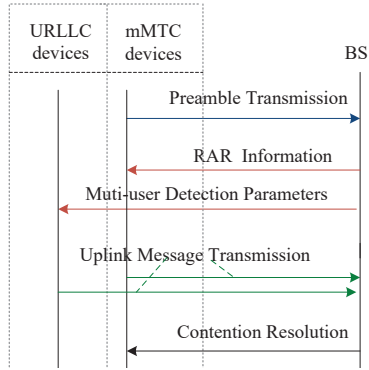}\\
	\caption{The proposed random access procedure.}\label{PR}
\end{figure}

\subsection{The LSTMH-RA Scheme Description}
Fig.~\ref{PR} illustrates the proposed LSTMH-RA scheme, which aims at meeting the diverse QoS requirements  in 6G  MTC networks. Specifically,  mMTC  devices access the network via  a timing advance (TA)-aided four-step procedure  to meet  massive access requirement,  while  the access procedure of the URLLC devices is completed in two steps  coupled with  the mMTC  devices' access procedure to  reduce latency. 
Furthermore,  we propose an attention-based LSTM prediction model to predict the number of active URLLC devices, thereby  determining   the parameters of the multi-user detection algorithm to  guarantee the latency and reliability access requirements of URLLC devices.
The details are described as follows.

\textbf{Step 1: Preamble Transmission}




Each active mMTC device randomly selects a preamble sequence from  the preambles pool $\bm{P_o=\{\phi_1,\cdots,\phi_t,\cdots,\phi_{\tau_p}}\}$  where $\bm{\phi_t}$ is a Zadoff-Chu (ZC)  sequence with length $L_\text{p}$~\cite{ZC}  and ${\tau_p}$ is the number of preambles. Note that,  to overcome the  propagation
delay, the cyclic prefix (CP) is usually added as the head of preamble sequence.



Let $\rho_k^{\text{p}}$ denote the  preamble transmitting power, and $b(k)$ be the index of preamble selected by device $k$. Let $\bm{h_k}=\sqrt{\beta_k}\bm{g_k} \in \mathbb{C}^{M\times1}$ denote the channel state information (CSI) between device $k$ and the BS, where $\beta_k$ denotes the  large scale fading
parameter which can be known for device $k$~\cite{SUCRIPA} and $\bm{g_k}$ is the small scale fading with each element in $\bm{g_k}$ having a distribution of $\mathcal{C}\mathcal{N}(0,1)$.


Then, the received preamble signal $\bm{Y}$  is given by
\begin{equation}\label{1}
\begin{array}{l}
 \bm{Y}=\sum\limits_{i=1}^{\zeta}\sum\limits_{k\in \mathcal{A}_i}\sqrt{\rho_k^{\text{p}}\beta_k}\bm{g_k}
 (\bm{\phi}_{b(k),i})^{T}+\bm{N},
\end{array}
\end{equation}
where $\bm{N}\in \mathbb{C}^{M \times L_p }$  stands for the  additive complex Gaussian white noise   matrix with each  element having a distribution of  $\mathcal{CN}(0,{{\sigma }^{2}})$, and $\bm{\phi}_{b(k),i}\in \mathbb{C}^{L_p \times 1}$ is obtained by cyclically shifted  $i-1$ symbols from  preamble sequence  $\bm{\phi}_{b(k)}$ due to the propagation
delay~\cite{xiaojie}. We utilize power control $\rho_k^{\text{p}}\beta_k=1$ to make all preambles have the same  received power. In addition, each  mMTC device should also  transmit its 
 identification (ID) information
to the BS, to facilitate the BS to detect the collision-free devices. 



\textbf{Step 2: Random Access Response (RAR) Transmission and Muti-User Detection Parameters Broadcasting}

Through performing the circular convolution operation on the received preamble signal $\bm{Y}$ and preamble sequence $\bm{\phi}_t$,  the $i^{th}$ element is given by~\cite{xiaojie}
\begin{equation}\label{a1}
\begin{array}{l}
\bm{y_{t,i}}=\left[\bm{Y}\otimes\frac{\bm{\phi}^{*}_{t}}{||\bm{\phi}_{t}||}\right]_i=\sqrt{L_\text{p}}\sum\limits_{m\in \mathcal{C}_{t,i}}\bm{g_m}
+\bm{N},
\end{array}
\end{equation}
where 
$\mathcal{C}_{t,i}$ denotes the set of devices selecting preamble $\bm{\phi_t}$ in the $i^{th}$ annulus. $|\mathcal{C}_{t,i}|=1$ means that the mMTC device selecting $\bm{\phi_t}$ in the $i^{th}$  annulus  is free  from TA  collision, and thus the BS can successfully decode the ID information of this mMTC device. In other words, the BS can identify mMTC devices with  TA collision-free. 
 Then, the BS generates a RAR for each TA collision-free mMTC device, mainly including preamble identification $t$, TA index, and resource blocks.



Then, the BS predicts the number of active URLLC devices via utilizing our proposed attention-based
 LSTM prediction  model, which is described in Section II-B. Given the predicted number of active URLLC devices and the channel environment, the BS determines the parameters of muti-user detection (including the resource block, modulation and code schemes), and  {\color{red}{the BS broadcasts a pilot  followed by the muti-user detection parameters via the  broadcast channel, where the broadcasted pilot is  known to all active devices.}}



\textbf{Step 3:  Uplink Message Transmission}


 Based on the received RARs,  each mMTC device first finds  RARs corresponding to its selected preamble,  and then matches the TA information in these RARs with its own. If the TA information contained in one RAR is the same as its own, it selects  the highest  power level $l_1$, and utilizes it  to  obtain  the transmitting power of  uplink message according to $\frac{l_1}{\beta_k}$.
 Otherwise, this mMTC device randomly selects a resource block from resource blocks corresponding to its selected preamble, randomly selects a  power level (denoted by $l_d, d=2,\ldots,L$), and transmits its uplink message with transmitting power $\frac{l_d}{\beta_k}$ via the selected resource block. {\color{red} Note that, the uplink message of each mMTC device is encoded by the low-density parity-check (LDPC) code according to the 3GPP specification~\cite{LDPC}.}

 In addition, {\color{red} using  the received pilot signal, each URLLC device  estimates its channel state information and thus obtains the muti-user detection parameters by correlating the received message with the estimated channel state information.} Then, based on the broadcasted parameters of muti-user detection, each URLLC device modulates and encodes its payload data to obtain its uplink message, and transmits it to the BS via the allocated resource block. 

\textbf{Step 4: Contention Resolution}

The BS utilizes the SIC algorithm in power domain to detect the received uplink message in each resource block allocated to  mMTC
devices. The SIC algorithm decodes the uplink messages of mMTC devices from the highest to the lowest power level.
The following two events ensure that a device selecting power level $l$ can be successfully decoded~\cite{pp}: 1) this device is free from power  collision  and  the messages from  devices with power levels  larger than  $l$  are successfully decoded; 2) this device is free from power collision  and  the number of devices with power levels  larger than  $l$ is zero. If the messages from this device can be decoded successfully,  the interference of the uplink message of this device will be cancelled from the received uplink message. Otherwise, this device fails to access the network and will access the network in the upcoming random access time slot.

 The BS decodes the uplink messages of URLLC devices via using the muti-user detection algorithm, which is not the focus of our paper and readers who are interested in   this  can refer to papers~\cite{mul1,mul2,mul3}~and references therein. 
 The coding and modulation schemes can be adjusted to satisfy the reliability requirement of URLLC devices based on the number of active URLLC devices~\cite{mul1,mul2,mul3}. 
 If the predicted number of URLLC devices is less than the actual value, we assume that all URLLC devices cannot be successfully decoded due to the incorrect coding and modulation schemes.
Section III-C will show that, the probability of such an event is really small. 
{\color{red}{Note that, if the BS successfully decodes the uplink message of one device, the BS will send ACK information to this device. Thus, the device can realize that its uplink message has been successfully detected.} 

In addition, in the proposed LSTMH-RA scheme, 
MTC devices access the  network  via  a  TA-aided  four-step procedure to meet massive access requirement, while the access  procedure  of  the  URLLC  devices  is  completed in  two  steps  coupled  with  the  mMTC  devices’  access procedure  to  reduce  latency.  The BS does not need to know the number of active mMTC devices for the access procedure of mMTC devices, and the 
  BS  predicts  the  number  of  active  URLLC devices via utilizing our proposed attention-based LSTM prediction  model for the access procedure of mMTC devices. Thus, the proposed LSTMH-RA scheme can be used in practice. Furthermore, for mMTC devices, the proposed access procedure consumes two round-trip process. From the procedure of the proposed scheme, we observe that, the delay depends largely on  step 4 of the proposed access procedure because the BS needs to decode the uplink message of active mMTC device during this step. Fortunately, the demodulation chip achieves lower demodulation delay. Hence, the proposed scheme  will cause a lower delay for mMTC devices. In addition, most mMTC devices is in low/no mobility~\cite{5G}. Due to the fact that  the  device’s speed  is inversely proportional to the coherence time, the proposed scheme can be used for mMTC devices.}                                                                                         
 {\color{red}\begin{remark}[What happens to the extremely overloaded network]
 In the  proposed LSTMH-RA scheme, if mMTC devices selecting the same preamble can be distinguished by utilizing the TA information, the BS will allocate resources to these  distinguished  TAs corresponding to these distinguished mMTC devices. Then, only distinguished mMTC devices  will utilize the highest  power level to transmit their uplink messages via the allocated resources corresponding to their TAs. By doing so, the BS utilizes the SIC algorithm to successfully decode the uplink messages of these distinguished mMTC devices.  Otherwise, the mMTC device  will randomly select a resource block from resource blocks corresponding to its selected preamble and randomly select a  power level to transmit its uplink message. 
Hence, when the network is extremely overloaded, the probability that mMTC devices selecting the same preamble but having different TA information will be very low, leading to the  increase of  power collision for the devices  whose TA information cannot be distinguished.  
 As a result, the throughput of the proposed LSTMH-RA scheme will be low. To solve this problem, the access class barring (ACB) mechanism~\cite{3gpp1} can be utilized  to reduce traffic overload as in~\cite{SUCRIPA}. More specifically, the BS broadcasts ACB factor  before step 1 of the proposed LSTMH-RA scheme. Then,  mMTC devices  passing the ACB check will perform the LSTMH-RA procedure. 
 \end{remark}}

\subsection{Attention-based LSTM  prediction model}

To ensure the reliability requirement  of URLLC device, our proposed random access scheme predicts the number of active URLLC devices during each random access time slot. {\color{red} Convolutional Neural Network (CNN) and  LSTM  are two generally used deep-learning models, where  CNN is mainly used to handle image data and  LSTM is usually used
for sequential data~\cite{hanDL}. The number of URLLC devices follows Poisson distribution, and can be considered as sequential inputs. In addition, attention mechanism is a useful tool  to improve the performance
of  the LSTM model~\cite{attention}.  Hence, we  use the attention-based LSTM prediction model.}

\begin{figure*}[htbp]
	\centering
	\includegraphics[scale =0.5] {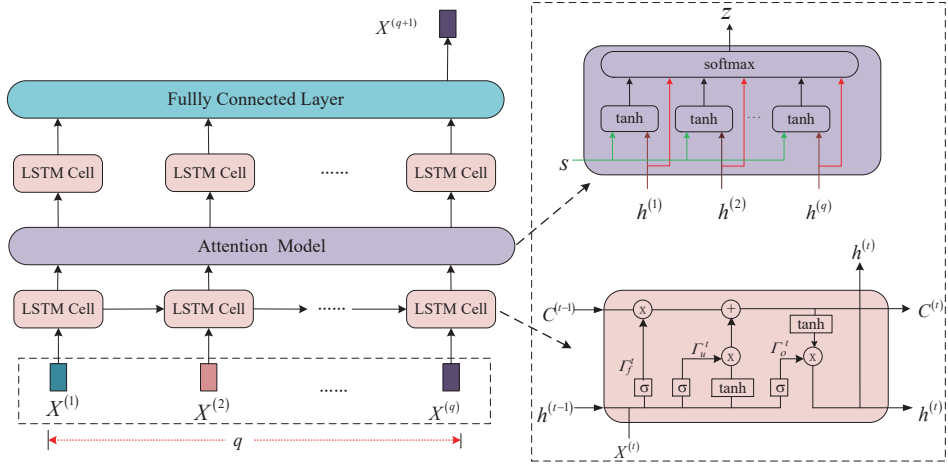}\\
	\caption{An attention-based LSTM prediction model consisting of  two LSTM layers, one attention layer and one fully connected layer.}\label{LSTM-model}
\end{figure*}

This prediction model consists of  two LSTM layers, one attention layer and one fully connected layer, as shown in Fig.~\ref{LSTM-model}.  The  data set are the input of the first LSTM layer, and the outputs of the first LSTM layer are fed into the attention layer. Then, we take the output of the attention layer as the input of the second LSTM layer, and the output of the second LSTM layer is connected to a fully connected layer to predict the number of active URLLC devices. Our training data is the number of URLLC devices from  $t+1$ to $t+q$, which are generated  based on the distribution model of URLLC devices. Furthermore, to ensure the reliability and latency requirements of URLLC devices, we take the maximum  number of active URLLC devices from $t$ to $t+d$ random access time slots  as the possible value in the $t^{th}$
random access time slot. In the following, we describe this prediction model in detail.

\subsubsection{LSTM layer}

The  powerful ability in preserving temporal memory makes LSTM inherently advantageous to perform the prediction task. Each LSTM layer consists of multiple LSTM cells. An LSTM cell consists of three kinds of gates, namely input gate, output gate, and forget gate. The input data of the first LSTM layer  is $\bm{X}=[X^{(1)},X^{(2)},\cdots,X^{(q)}]$, where $q$ is the length of input. The process of building gates  during the $t^{th}$ time step is given by~\cite{2016LSTM}
\begin{equation}\label{lstm}
\begin{array}{l}
\left\{
\begin{array}{l}
\Gamma_{f}^{(t)}=\delta(W^{(f)}[h^{(t-1)},X^{(t)}]+b^{(f)}) \\
\Gamma_{u}^{(t)}=\delta(W^{(u)}[h^{(t-1)},X^{(t)}]+b^{(u)})\\
{\tilde{C}}^{(t)}=\mathrm{tanh}(W^{(c)}[h^{(t-1)},X^{(t)}]+(b^{(c)})\\
{C}^{(t)}=\Gamma_{f}^{t}\cdot C^{(t-1)}+\Gamma_{u}^{(t)}\cdot \tilde{C}^{(t)}\\
\Gamma_{o}^{(t)}=\delta(W^{(o)}[h^{(t-1)},X^{(t)}]+b^{(o)})\\
h^{(t)}=\Gamma_{o}^{(t)}\cdot \mathrm{tanh}(C^{(t)}),
\end{array}
\right.
\end{array}
\end{equation}
where $\Gamma_{f}^{(t)}$, ${C}^{(t)}$ and $h^{(t)}$ denote the forget gate, the input gate, and the output gate, respectively, while $W$ and $b$ represent the weight and the bias factors of different structures, respectively. Furthermore, $\delta$ denotes $\mathrm{sigmoid}$ activation function.

\subsubsection{Attention layer}
Attention model (AM) was first introduced from the machine translation task and has become a popular  neural network concept~\cite{2014Neural}.
Attention can integrate related information and allow the model to provide dynamic attention to some useful input information, which has been utilized as a useful tool to improve the performance of LSTM network.


We assume that the output of the $j^{th}$ time step in the $i^{th}$ layer is ${h_{j}^{i}}~(i=1,2,~j=1,\ldots, q)$
\begin{equation}\label{Ag}
z^{i} = \sum\limits_{j=1}^{q} {a_{j}^{i}h_{j}^{i-1}},
\end{equation}
 where $a_{j}^{i}$ is the attention weight. It can be given by
 \begin{equation}\label{Ag1}
a_{j}^{i} = \frac{{\mathrm{exp} (e_{j}^{i})}}{{\sum\nolimits_{j = 1}^{q} {\mathrm{exp} (e_{j}^{i})} }},
\end{equation}
 where $e_{j}^{i}$ is  the aggregation state. $e_{j}^{i}$ can be  obtained via utilizing the Bahdanau-attention method~\cite{2014Neural}
\begin{equation}\label{Ag2}
e_{j}^{i}= {V^T}\mathrm{tanh} ({W_u}\bm{S} + {W_h}h_{j-1}^{i}),
\end{equation}
where $V, W_u$ and ${W_h}$ are the input weights, and $\bm{S}=[{h_{1}^{i-1}},\ldots, {h_{q}^{i-1}}] $ is  the output of the ${(i-1)^{th}}$ LSTM layer.

Finally, we use the combination of the attention model output $z^{i}$ and the LSTM output $[{h_{1}^{i-1}},\ldots, {h_{q}^{i-1}}]$ as the input data for the  second LSTM layer, i.e.,  $\mathrm{concat}(z^{i},{h_{1}^{i-1}},\ldots, {h_{q}^{i-1}})$.


\subsubsection{Output layer}
 The extracted features in the last LSTM layer go through a fully connected layer to obtain the output  $X^{(q+1)}$.  Note that, the output $X^{(q+1)}$ may not be an integer, and we should  round $X^{(q+1)}$ after prediction.

\section{System analysis}

The proposed random access scheme aims to satisfy different communication requirements of URLLC and mMTC devices, i.e., high reliability and low
access latency for URLLC devices and  massive access for mMTC devices.

Based on the predicted  number of active URLLC devices, the  BS determines  the  parameters  of  the  multi-user  detection algorithm to ensure the reliability requirement. The coding and modulation schemes can be adjusted to satisfy the reliability requirement of URLLC devices based on the number of active URLLC devices~\cite{mul1,mul2,mul3}. 
 If the predicted number of URLLC devices is less than the actual value, we assume that all URLLC devices cannot be successfully decoded due to the incorrect coding and modulation schemes. Fig.~{\ref{error}}  will show that the probability of such an event, denoted by $P_\text{LSTM}$, is really small. Thus, the proposed random access scheme can ensure the  reliability and 
access latency requirements of URLLC devices. Therefore, in this section,  we only focus on the number of successful access devices, which is given by
 \begin{equation}\label{2-18}
\begin{array}{l}
N^{s}=N_\text{U}^{s}+N_\text{M}^{s},
\end{array}
\end{equation}
where $N_\text{U}^{s}$ and $N_\text{M}^{s}$ are the number of successful URLLC devices  and mMTC devices, respectively.


For the active URLLC devices, we assume that, if  the  predicted  number  of active URLLC  devices is  less  than  the  real  value,  all active  URLLC devices  cannot  be  successfully  decoded. Then,  we have
 \begin{equation}\label{2-18U}
N_\text{U}^{s}=N_\text{U}(1-P_\text{LSTM}).
\end{equation}
It is hard to derive the  expression of $P_\text{LSTM}$ since the number  of  active URLLC  devices is predicted by  an attention-based LSTM prediction model. However, we can compute the expected value  though  Monte Carlo simulations.
Next, we discuss how to derive the term $N_\text{M}^{s}$ in (\ref{2-18}).

Let $P_{N_\text{M}}^u$ denote the probability of $u$ devices selecting the same preamble among ${N_\text{M}}$ devices. $u$  follows a binomial distribution  ${{u}}\sim{}\mathcal{B}(N_\text{M},\frac{1}{{{\tau }_\text{M}}})$,  and  $ P_{N_\text{M}}^u$ is given by
\begin{equation}
   P_{N_\text{M}}^u
= \left( \begin{array}{l}
{N_\text{M}}\\
u
\end{array} \right){\left( {\frac{1}{{{\tau_p}}}} \right)^u}{\left( {1 - \frac{1}{{{\tau_p}}}} \right)^{{N_\text{M}} - u}}. 
\end{equation}

Let $P_u^{t}$ denote the probability that $t$ resource blocks are allocated to these $u$ devices, i.e., $t$ devices experience no  TA  collision among these $u$ devices. Let $N_{u,t}^{s}$ denote the number of successful devices when $u$ devices  are allocated $t$
 resource blocks.
Then,  $N_\text{M}^{s}$ is computed by
\begin{equation}\label{2-19}
\begin{array}{l}
N_\text{M}^{s}={\tau }_\text{M}\sum\limits_{u = 1}^{{N_{{\rm{M}}}}} {\sum\limits_{t = 1(t \ne u - 1)}^u {N_{u,t}^sP_u^tP_{{N_{{\rm{M}}}}}^u} }.
\end{array}
\end{equation}

Next, we present how to compute  terms $P_u^t$ and $N_{u,t}^s$ in (\ref{2-19}) in detail.

{$ {{(1)}}$ $P_u^t$ computation}

$P_u^{t}$ denotes the probability that $t$ resource blocks can be allocated to  $u$ devices.
This means that,  $t$ devices have unique TA values out of  $u$ devices, and, among the remaining $u-t$ devices, each device has the same TA value as at least one other device.
Therefore,  to compute the value of $P_u^{t}$, we first  need to
find cases of each device has the same TA value as at least one other device among the remaining $u-t$ devices,  and then derive TA values allocation ways for each case. The details are described as follows.

 Finding such cases is equivalent to finding  sets  where each element (represents  the number of UEs having the same TA value) is larger than 1 and the sum of elements  equals $u-t$, which is given by
\begin{equation}\label{S}
 \begin{array}{l}
{\rm{Find }}~\bm{S} = [\bm{S}(1),\cdots,\bm{S}(j),\cdots,\bm{S}(|\bm{S}|)]\\
{\rm{s.t.  }}~\bm{S}(i) > 1,\\
{\rm{       }}~~\sum\limits_{i = 1}^{|\bm{S}|} {\bm{S}(i)}  = u - t,
\end{array}
\end{equation}
where $\bm{S}(j)$ denotes the $j^{th}$ element in  set $\bm{S}$, and $|\bm{S}|$  represents the number of different TA values.

Let $N_{u-t}$ denote the number of sets  satisfying  constraints in (\ref{S}), and  $\bm{S_{u-t}^i}$  denote the $i^{th}$ set.
For the case of ${u-t}=1$, it is easy to derive that $\bm{S_1^i}$ is an empty set; for the case of ${u-t}=2$ , it is easy to derive that $N_{2}=1$ and $\bm{S_2^1}=[2]$; for the case of ${u-t}=3$, it is easy to derive that $N_{u-t}=1$ and $\bm{S_3^1}=[3]$; for the case of ${u-t}\ge 4 $,  sets satisfying  constraints in (\ref{S}) are derived via  a recursive method, which is shown in Algorithm 1.
 \begin{algorithm}[!htbp]
	\SetKwInOut{KIN}{Input}
	\SetKwInOut{KOUT}{Output}
	\SetKwInOut{Return}{Return}
	\caption{Recursive method of deriving  sets satisfying  constraints in (\ref{S}) when $u-t \ge 4$ }
	\KIN { $N_2, N_3, [\bm{S_{2}^1},\ldots, \bm{S_{2}^{N_2}}],[\bm{S_{3}^1},\ldots, \bm{S_{3}^{N_3}}]$}
	\KOUT {$\mathcal{D}=[\bm{S_{u-t}^1},\bm{S_{u-t}^2}, \cdots, \bm{S_{u-t}^{N_{u-t}}}],{N_{u-t}}$}
		$\bm{\mathcal{C}}=\phi, \bm{\mathcal{D}}=\phi$, $\bm{S_{u-t}^1}=[r]$\;
$\bm{\mathcal{C}}=\bm{\mathcal{C}} \cup \bm{S_{u-t}^1}$\;
	\For{  all $a \in \{2,\ldots, \left\lceil {r/2} \right\rceil\}$ }
		{			
			$b=(u-t)-a$\;
          $\bm{\mathcal{A}}=[\bm{S_a^1},\cdots,\bm{S_a^{N_{a}}}]$, $\bm{\mathcal{B}}=[\bm{S_b^1},\cdots,\bm{S_b^{N_{b}}}]$\;
          \textbf{Combine each element in} $\bm{\mathcal{A}}$ \textbf{with each element in} $\bm{\mathcal{B}}$\;
          \For{  all $i\in \{1,\ldots, {N_{a}}\},  j\in \{1,\ldots, {N_{b}}\}$ }
          {
            $\bm{\mathcal{C}}=\bm{\mathcal{C}} \cup [\bm{S_a^i},\bm{S_b^j}]$\;
            }

		}
$\bm{\mathcal{D}}=[\bm{S_{u-t}^1},\bm{S_{u-t}^2}, \cdots, \bm{S_{u-t}^{N_{u-t}}}]\leftarrow$ Regard sets in $\bm{\mathcal{C}}$ having the same elements but in different order  as one set\;
	\Return{
$\mathcal{D}=[\bm{S_{u-t}^1},\cdots, \bm{S_{u-t}^{N_{u-t}}}], {N_{u-t}}=|\mathcal{D}|$}
\end{algorithm}

For the $i^{th}$  set $\bm{S_{u-t}^i}$,   since  $t$
devices have unique TA values and the remaining $u-t$ devices  have   $|\bm{\bm{S_{u-t}^i}}|$ different TA values   among $u$ devices,  the total number of different TA values among these $u$ devices is $N_\text{TA}^{i}=|\bm{\bm{S_{u-t}^i}}|+t$. Since there are $\zeta$ different TA values in  the cell (i.e., $\text{TA}_1,\cdots,\text{TA}_\zeta$), there are $N_\text{TAC}^{i}=\binom{\zeta}{{N_\text{TA}^{i}}}{N_\text{TA}^{i}}! $ different TA allocation ways if $\zeta \ge {N_\text{TA}^{i}}$.  Otherwise, $N_\text{TAC}^{i}=0$.
Then, we have
\begin{equation}\label{2-20}
\begin{array}{l}
P_u^t=\sum\limits_{i= 1,{N_\text{TAC}^{i}}\ge 1 }^{N_{u-t}}{\sum\limits_{k= 1}^{N_\text{TAC}^{i}} {N_\text{do}}
\prod\limits_{g = 1}^{N_\text{TA}^{i}}(P_{k_g^{i}})^{\bm{\bm{S_{u-t}^i}}(g)}\prod\limits_{c =N_\text{TA}^{i}+1}^{N_\text{TA}^{i}+t}P_{k_c^{i}}},
\end{array}
\end{equation}
where $P_{k_g^{i}}$ is the probability that the TA index of a device is ${k_g^{i}}$ (i.e., the probability that a device is located at the $({k_g^{i}})^{th}$ annulus),  which is given by
\begin{equation}
\begin{array}{rcl}
{P_{k_g^{i}}}=\left\{
\begin{aligned}
\int_{\frac{d({k_g^{i}}-1)}{2}}^{\frac{2d{k_g^{i}}}{2}}\dfrac{2x}{{R}^2} {dx}&=\dfrac {d^2{(2{k_g^{i}}-1)}} {4{R}^2},\\
 &{{{k_g^{i}}=1,\ldots,{\zeta-1}}},\\
\int_{\frac{d({k_g^{i}}-1)}{2}}^{R} \dfrac{2x}{{R}^2} {dx}&=1-\dfrac {d^2{({k_g^{i}}-1)^2}} {4{R}^2},\\
&{{k_g^{i}}=\zeta},
\end{aligned}
\right.
\end{array}
\end{equation}
and ${N_\text{do}}$ stands for the order of these $u$ devices, which is given by
\begin{equation}\label{2-21}
\begin{array}{l}
{N_\text{do}}=\frac{\binom{u}{t} \binom{u-t}{\bm{S_{u-t}^i}(1)}
\prod\limits_{d = 2}^{|\bm{S_i}|}\binom{u-\sum\limits_{r=1}^{d-1}\bm{\bm{S_{u-t}^i}}(r)}{\bm{\bm{S_{u-t}^i}}(d)}}{\prod\limits_{j = 1}^{|\bm{D_i}|}c_i^{j}!},
\end{array}
\end{equation}
where $\bm{D_i}$ represents a set consisting of different elements in $\bm{S_{u-t}^i}$ and $c_i^{j},~j=1,\ldots, |\bm{D_i}|$  denotes the number of times that the $j^{th}$ element of $\bm{D_i}$ appears in set $\bm{S_{u-t}^i}$.

{$ {{(2)}}$ $N_{u,t}^{s}$  computation}

Based on the procedure of our proposed random access scheme, $u$ devices selecting the same  preamble are allocated $t$
 resource blocks. On each resource block, only one device has the highest power level. Thus,  these $t$ devices can successfully access the network based on the SIC algorithm in power domain. The remaining $u-t$ devices first randomly select their resource blocks from these $t$ resource blocks, and then randomly select their power levels from $L-1$ available power levels. Let $N_{u-t}^{s}$  denote the number of successful devices among the remaining $u-t$. Then, $N_{u,t}^{s}$ can be written as
 \begin{equation}\label{ts0}
N_{u,t}^{s} =t+N_{u-t}^s.
\end{equation}
Next,  we discuss how to derive $N_{u-t}^{s}$.

  Let $P_i$ denote the probability of $i$ devices selecting the same resource block  among these $u-t$ devices. Apparently, $i$  follows a binomial distribution of ${{i}}\sim{}\mathcal{B}(u-t,\frac{1}{t})$, i.e., $P_i=\left( \begin{array}{l}
u - t\\
i
\end{array} \right){\left( {\frac{1}{t}} \right)^i}{\left( {1 - \frac{1}{t}} \right)^{u - t - i}}$.  Let $ P_i^s$  denote the probability that one device can be successfully decoded via the SIC algorithm for the remaining $u-t$ devices.  Then, we have
\begin{equation}\label{ts}
  N_{u-t}^s= \sum\limits_{i = 1}^{u - t}{i\times t \times P_i^s\times P_i}.
\end{equation}
Based on the principle  of the SIC algorithm in power domain, in each resource block,  the following two events ensure that  device selecting power level $l$ can be successfully decoded: 1) this device is free from power level collision  and  devices with power levels  larger than this device are successfully decoded; 2) this device is free from power level collision  and  the number of devices with power levels  larger than this device is zero. Let $N_{\text{E1},l}^s$ and $N_{\text{E2},l}^s$  denote the cases of the first event and the second event, respectively.   Apparently, $N_{\text{E2},l}^s={{\left( {L - 2} \right)}^{i - 1}}$  and the total number of cases selecting power level is  ${{(L-1)^i}}$.  Then, $ P_i^s$ is given by
\begin{equation}\label{ps}
 \begin{array}{l}
 P_i^s = \dfrac{{\sum\limits_{l = 1}^
 {L-1}{\left[ {N_{\text{E1},l}^s{\rm{ + }}N_{\text{E2},l}^s} \right]} }}{{{L^i}}}\\ \\
 = \frac{{\sum\limits_{l = 1}^{L-1} {\left[ {\sum\limits_{j = 1}^{l - 1} {\left( \begin{array}{l}
i - 1\\
j
\end{array} \right)\left( \begin{array}{l}
l - 1\\
j
\end{array} \right)j!{{\left( {L -1-l} \right)}^{i - j - 1}} + {{\left( {L - 1-l} \right)}^{i - 1}}} } \right]} }}{{{(L-1)^i}}}.
\end{array}
\end{equation}

\section{Numerical Results}

In the following,   we first present the  performance of the attention-based LSTM  prediction model, and then make a comparison  between our proposed  LSTMH-RA scheme and the random access proposed in~\cite{co1} in  terms of  the number of successful access devices.

\begin{table}[htbp]
\scriptsize
\centering
 \caption{Simulation parameters}\label{tab:parameters}
   \begin{tabular}{|c|c|}
\hline
\linespread{2}
Parameter & Value\\
\hline
$\lambda$~ (Mean value of Poisson  distribution) & 4$\sim$8\\
\hline
$d$~(\tabincell{c}{The  quantized  unit })& 156 $m$\\
\hline
$T_s$~(the basic time unit)&  $3.072\times 10^{-7}~s$\\
\hline
$c$~(The speed of light)& $3\times10^8m/s$\\
\hline
$R$~(The radius of cell)& 624, 1000, 1500 $m$ \\
\hline
 $\tau_p$~ \tabincell{c}{The number of preambles}& 28 $\sim$ 64\\
\hline
$N_\text{M}$~(The number of active mMTC devices) &  100\\
\hline
$L$~(The number of power levels)& 3 $\sim$ 7\\
\hline
$M$~(The number of antennas)   & 128\\
\hline
$N_a$~\tabincell{c}{(The number of active UEs during \\ one random access procedure)} & 15 $\sim$ 50\\
\hline
$N_\text{U}$~(The number of active URLLC devices) & 5\\
\hline
\end{tabular}
\end{table}

\subsection{The performance of the attention-based LSTM  prediction model}


 For  the attention-based LSTM prediction model, we set the learning rate to 0.001, and utilize the root-mean square loss function. As in~\cite{possion1}, we assume that the number of active URLLC devices in each random access time slot has a distribution of  $\mathcal{P}(\lambda)$ where $\lambda$ is the mean value.

\begin{figure}[htbp]
	\centering
	\includegraphics[scale=0.5]{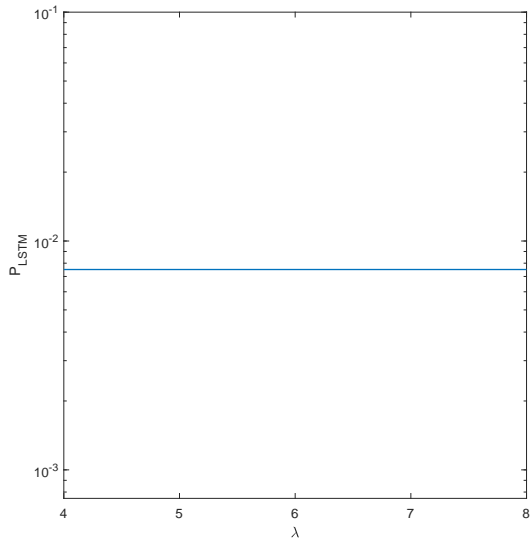}\\
	\caption{The probability of the predicted number of active URLLC being less than the actual value.}\label{error}
\end{figure}
\begin{figure}[htbp]
	\centering
	\includegraphics[scale=0.5]{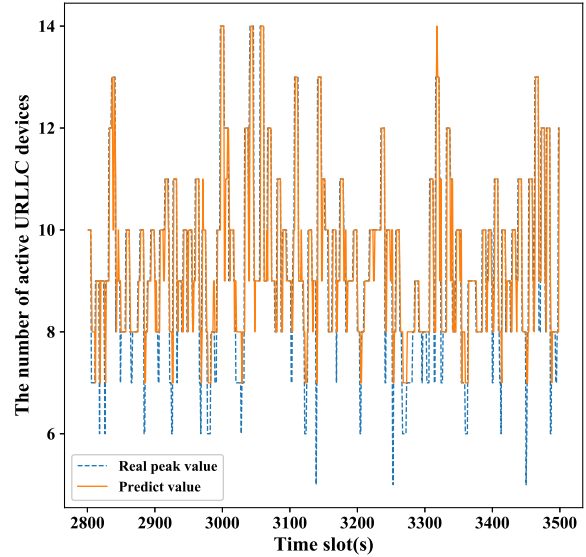}\\
	\caption{Prediction results.}\label{LSTMP}
\end{figure}
\begin{figure}[htbp]
	\centering
	\includegraphics[scale=0.5]{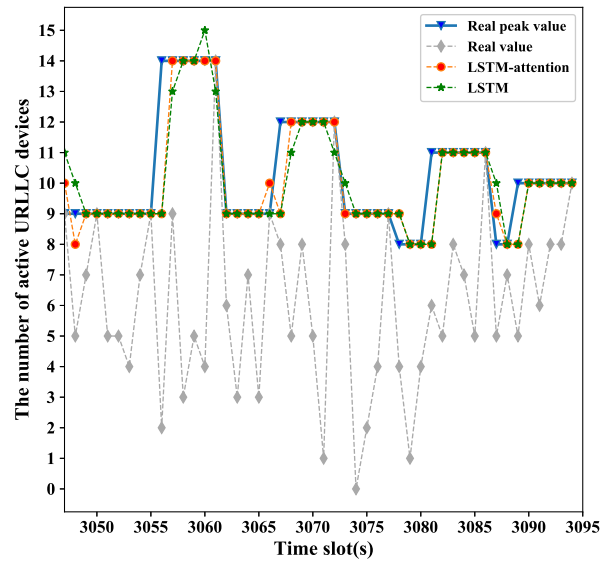}\\
	\caption{Prediction comparison.}\label{LSTMnn}
\end{figure}

If the predicted number of URLLC devices is less than the actual value, we assume that all URLLC devices cannot be successfully decoded due to the incorrect coding and modulation schemes. To  ensure the reliability and latency requirements of URLLC devices, we take the maximum  number of active URLLC devices from $t$ to $t+5$ random access time slots  as the possible value in the $t^{th}$
random access time slot, and set the length of input be equal to $q=5$.

{\color{red}Fig.~\ref{error} shows how the probability $P_\text{LSTM}$  changes with  different  arrival rates $\lambda$ ranging from 4 to 8. We see from this figure that,  $P_\text{LSTM}$ takes small  and almost the same values for different  arrival rates $\lambda$.} This means that our proposed prediction model can  ensure the reliability and latency requirements of URLLC devices with high probability. 
This indicates that the allocated code and modulation parameters can ensure the reliable communication of almost all active URLLC devices.
In addition, for the case where the  predicted  number  of  URLLC  devices is  less  than  the  actual value, the BS will allocate multiple resource blocks to these  active URLLC devices in the next random access time slot to ensure all  active URLLC devices can be decoded successfully within two random access time  slots to satisfy the access latency requirement. Thus, the proposed random access scheme can ensure the  reliability and 
access latency requirements of URLLC devices.


Fig.~\ref{LSTMP} shows the prediction results when we consider multiple random access time slots, where we set $\lambda$ to 6. The `real peak value' in this figure refers to  the maximum number of active URLLC devices from $t$ to $t+5$ random access time slots.
We can see from this figure that, the predicted value is very close to the actual value, which indicates that the prediction model is efficient to predict the maximum  number of active URLLC devices from $t$ to $t+5$ random access time slots. In addition, for the case where the  predicted  number  of  URLLC  devices is  less  than  the  actual value, the BS will allocate multiple resource blocks to these  active URLLC devices in the next random access time slot to ensure all  active URLLC devices can be decoded successfully within two random access time  slots to satisfy the access latency requirement. Thus, the proposed random access scheme can ensure the  reliability and  access latency requirements of URLLC devices.

Fig.~\ref{LSTMnn} compares the prediction performance between the proposed attention-based LSTM and basic LSTM prediction model, where we set $\lambda$ to 6.   The LSTM prediction model is obtained by deleting the attention model from our attention-based LSTM model. Furthermore,   `real value' is the number of URLLC devices in the $t^{th}$ random access time slot.
The experimental results show that, compared to the  basic LSTM prediction model, our proposed attention-based LSTM prediction model  can achieve better performance.

\subsection{The performance of the proposed LSTMRH-RA}

 In the simulations, we consider the urban micro scenario and all devices are uniformly distributed in the cell.  In addition, the quantized unit is $d=16T_sc=156~m$ where $T_s=3.072\times 10^{-7}s$ is the basic time unit~\cite{zhang}~and $c=3\times10^8m/s$ is the speed of light.

\begin{figure}[htbp]
	\centering
	\includegraphics[scale=0.4]{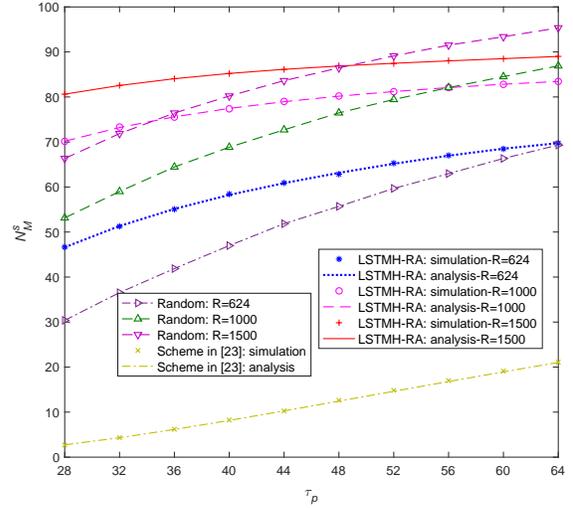}\\
	\caption{ The number of successful mMTC devices versus the number of preambles.}\label{TMtaop}
\end{figure}

Fig.~\ref{TMtaop} shows the number of successful mMTC devices with respect to the number of preambles for different radius of cell. We set the radius of cell to 1500, 1000, and 624, the number of active mMTC devices to $100$,  power levels to 3, and the number of antennas at the BS to 128.  {\color{red} In addition,  to demonstrate the advantage of the  power allocation strategy utilized in the proposed LSTMH-RA scheme, we  consider a  scheme which allows
 each device  to select its uplink transmit power randomly and uniformly. This scheme  is termed `Random' in Fig.~\ref{TMtaop}.  We observe  that, for different radius,  with the decrease of the number of preambles (i.e., the number of devices selecting the same preamble increasing), the gap between the proposed LSTMH-RA scheme  and the Random scheme increases. This indicates that the proposed LSTMH-RA scheme is more robust to the Random scheme for the crowded scenarios.} We also see from Fig.~\ref{TMtaop}  that, with the increase of the number of preambles, the number of successful mMTC devices increases and is significantly higher than that of the random access scheme proposed in~\cite{co1}. Furthermore, with the increase of the size of the  cell,  the number of successful mMTC devices increases. The reason is that, with the increase of the size of the cell, the number of different TA values increases  and thus the BS can distinguish more mMTC devices.

\begin{figure}[htbp]
	\centering
	\includegraphics[scale=0.45]{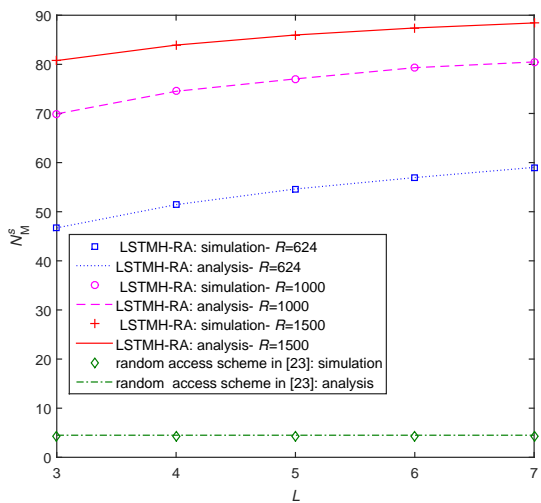}\\
	\caption{The number of successful mMTC devices versus the number of power levels.}\label{TML}
\end{figure}

Fig.~\ref{TML} shows  the number of successful mMTC devices with respect to the number of power levels for different radius of cell. We set the radius of cell to 1500, 1000, and 624. We set the number of active mMTC devices to 100, the number of preambles to $28$, and the number of antennas at the BS to 128.  We can see from Fig.~\ref{TML}  that, with the increase of the number  of power levels, the number of successful mMTC devices increases and is significantly higher than that of the random access scheme proposed in~\cite{co1}. Furthermore, with the increase of the size of the  cell,  the number of successful mMTC devices increases. The reason is the same as we described in Fig.~\ref{TMtaop}.
\begin{figure}[htbp]
	\centering
	\includegraphics[scale=0.45]{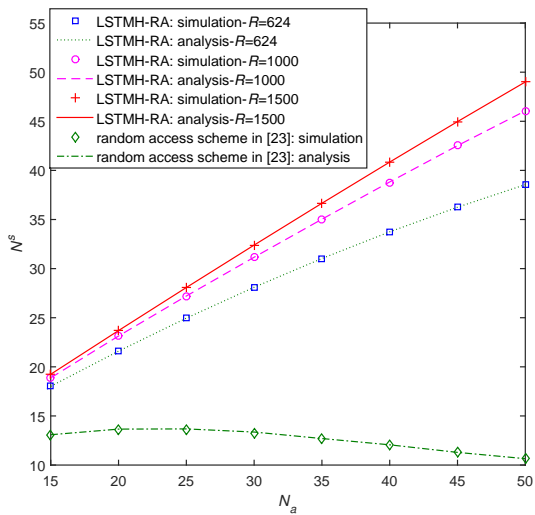}\\
	\caption{ The number of successful  devices versus the number of active devices.}\label{TN}
\end{figure}

Fig.~\ref{TN} shows how the number of successful  devices changes with the number of active devices.  We set the radius of cell to 1500, 1000, and 624,  the number of preambles to 28,  the number of antennas at the BS to 128, and the number of power levels to 3. Among these active devices, the number of active URLLC devices is set to 5 and these active URLLC devices belong to different groups, and thus the number of preambles allocated to the URLLC devices is 23 in the random access scheme proposed in~\cite{co1}. Based on~Fig.~{\ref{error}}, we can see that $P_\text{LSTM}$ in (\ref{2-18U}) is about $7.5\times10^{-3}$.
 We can see from Fig.~\ref{TN}  that, with the increase of the number of active devices, the number of successful devices increases and is significantly higher than that of the random access scheme proposed in~\cite{co1}. Furthermore, with the increase of the size of the  cell,  the number of successful mMTC devices increases. The reason is the same as we described in Fig.~\ref{TMtaop}.

\section{Conclusion}

In this paper, an  LSTMH-RA scheme has been proposed to support diverse QoS requirements in 6G MTC
networks where URLLC and mMTC devices coexist. To meet URLLC devices' latency and reliability access requirements, this
scheme employs a proposed attention-based LSTM prediction model to predict the number of active URLLC devices, and thus URLLC devices can access the
network via a two-step contention-free access procedure. In addition, to meet mMTC devices' massive access
requirement, mMTC devices access the network via a contention-based
TA-aided access mechanism. Numerical results have shown that,
compared with the existing schemes, the LSTMH-RA scheme
significantly improves the successful access probability,
and satisfies the diverse QoS requirements of URLLC and
mMTC devices.

\bibliographystyle{IEEEtran}
\bibliography{ref}
\end{document}